\documentclass[12pt]{article}
\usepackage{epsfig,graphicx}

\title{Theory  of hadron decay into baryon-antibaryon final state}
\author{ Yu.A.Simonov\\
State Research
Center\\Institute of Theoretical and Experimental Physics, \\
Moscow, 117218 Russia}

\newcommand{\beq}{\begin{eqnarray}}
 \newcommand{\eeq}{\end{eqnarray}}
\newcommand{\be}{\begin{equation}}
 \newcommand{\ee}{\end{equation}}

 \def\la{\mathrel{\mathpalette\fun <}}

\def\fun#1#2{\lower3.6pt\vbox{\baselineskip0pt\lineskip.9pt
\ialign{$\mathsurround=0pt#1\hfil ##\hfil$\crcr#2\crcr\sim\crcr}}}

\newcommand{{\SD}}{\rm SD}

\newcommand{\BB}{\mbox{${\mathcal{B}    \bar{\mathcal{B}}}$}}

\newcommand{\vex}{\mbox{\boldmath${\rm x}$}}
\newcommand{\vey}{\mbox{\boldmath${\rm y}$}}
\newcommand{\ver}{\mbox{\boldmath${\rm r}$}}
\newcommand{\vesig}{\mbox{\boldmath${\rm \sigma}$}}

\newcommand{\veP}{\mbox{\boldmath${\rm P}$}}
\newcommand{\vep}{\mbox{\boldmath${\rm p}$}}
\newcommand{\veq}{\mbox{\boldmath${\rm q}$}}

\newcommand{\veQ}{\mbox{\boldmath${\rm Q}$}}

\newcommand{\veR}{\mbox{\boldmath${\rm R}$}}

\newcommand{\veu}{\mbox{\boldmath${\rm u}$}}
\newcommand{\vev}{\mbox{\boldmath${\rm v}$}}

\newcommand{\vexi}{\mbox{\boldmath${\rm \xi}$}}
\newcommand{\veta}{\mbox{\boldmath${\rm \eta}$}}

\newcommand{\veal}{\mbox{\boldmath${\rm \alpha}$}}

\newcommand{\lan}{\langle}
\newcommand{\ran}{\rangle}

\begin{document}
\maketitle
\begin{abstract}

The  nonperturbative mechanism of baryon-antibaryon  production  due to  double
quark pair $(q\bar q)$($q\bar q)$ generation inside a hadron is  considered and
the amplitude is  calculated as matrix element of the vertex operator between
initial and final  hadron wave  functions. The vertex operator is expressed
solely in terms of first principle input: current quark masses, string tension
$\sigma$ and  $\alpha_s$. In contrast to meson-meson production via  single
pair generation, in baryon case  a new entity appears in the vertex: the vacuum
correlation length $\lambda $, which was computed  before through  string
tension $\sigma$. As an application electroproduction of $\Lambda_c
\bar\Lambda_c$ was calculated and an enhancement near 4.61 GeV was found in
agreement with recent experimental data.

\end{abstract}

\section{Introduction}
The  baryon-antibaryon $(\mathcal{B}\bar \mathcal{B})$ final states in hadron
reactions  are rather typical phenomena, e.g. in charmonium decays    $p\bar
p$, $\Lambda\bar \Lambda$ etc. channels are significant \cite{1}.  In $e^+e^-$
collisions the $\BB$ final states are carefully studied and display in many
cases $(\Lambda^+_c \Lambda^-_c, \Lambda\bar \Lambda, p\bar p)$ a  nontrivial
behavior near the corresponding thresholds, see \cite{1*} for a review and
references. In $B$ decays the produced $p\bar p$ pairs were observed with
near-threshold enhancements \cite{2}. In this paper we consider a rather
general type of reactions, when a quarkonia state  $(Q\bar Q)$ decays into
$\mathcal{B}\bar \mathcal{B}$, where $\mathcal{B}(\bar \mathcal{B}) $ contain
 quark $Q$  (antiquark $\bar Q$).
 From dynamical point of view, the
simplest case is the OZI allowed
 decay of heavy quarkonium into $\mathcal{B}\bar \mathcal{B}$ pair of  heavy-flavor baryons, e.g.
 $\psi (nS) \to \Lambda^+_c \Lambda_c^- $, which was experimentally observed
 first in \cite{3} at one energy, and  measured in \cite{4} in the mass
 interval $[4.5\div 5.4]$ GeV/$c^2$. For
this type of reaction the  creation of two light quark pairs is necessary and
one could expect some suppression in this channel. However, experimentally the
suppression is quite mild, as was discovered in the reaction  $e^+e^-\to
\Lambda^+_c \Lambda_c^-  $ in \cite{3,4}.

 A peak at the $\Lambda^+_c \Lambda^-_c$ mass around 4.63 GeV/$c^2$ was found
 in \cite{4}, and the nature of this enhancement is still obscure, however
 different explanations  were suggested \cite{5,6}.  A discussion
 of possible mechanisms of similar phenomena in  $\mathcal{B}\bar \mathcal{B} $
 produced in $B$ meson decays,  was given in \cite{7}.
 Below we develop a
 systematic theory of $\mathcal{B}\bar \mathcal{B}$  production in OZI allowed processes, which is
 actually a theory of  double string breaking  with $\mathcal{B}\bar \mathcal{B}$  emission, as shown in
 Fig.1. As it will be seen, this theory is a one-step development of the
 general approach of string breaking, given in \cite{8}. To simplify matter we
 consider  first the case of heavy quarkonium, decaying into heavy-flavor $\mathcal{B}\bar \mathcal{B}$
 pair.

\begin{figure}\vspace*{-4cm}
\begin{center}\hspace*{-8cm}
\includegraphics[height=25cm]{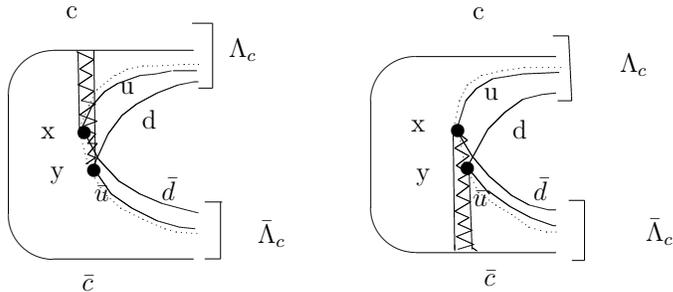}
\vspace*{-18cm}
 \center{ \caption{  Double quark-pair creation at points $x,y$ in
the field of heavy quarks $c\bar c$.} }\end{center}
\end{figure}

 \section{The formalism}

 The initial state of our problem is the heavy $Q\bar Q$ state, where $Q$ and
 $\bar Q$ are connected by a string. We are looking for a process, where two
 light $q\bar q$ pairs are created in the field of $Q\bar Q$, and hence the
 basic vertex is the $4q$ operator in the static $Q\bar Q$ confining field. As
 in the case of one pair  $q\bar q$ vertex, it is sufficient to consider the
 light quark Lagrangian in the field of the static antiquark $\bar Q$ and static quark
 $Q$.

 This  situation is shown in Fig.1, where a pair $\bar u u$ is created at the
 point $(\vex, x_4)$ and $\bar d d$ at $(\vey, y_4)$ with time growing from
 left to right. The  string junction trajectory is  shown in Fig. 1  by dotted lines and the
 string junction positions at each moment of time is defined as the Torricelli
 points in the  triangles formed by space positions of ($cud)$ and ($\bar c
 \bar u\bar d)$.

 It is important, that points $(\vex, x_4)$ and $(\vey, y_4)$ will be shown to
 be close to each other, and the string junction and anti-string junction are
 generated at one point in their vicinity,  which considerably facilitates the
 picture of $\BB$ creation (while the latter is rather complicated in a two-step
 $\BB$  production).

  We start with the partition function of a light quark in the  field of
  external current of heavy quarks $Q\bar Q$.

  \be Z = \int DA D\bar\psi D\psi \exp \left[ - (S_0 + S_1 + S_{int} + S_{Q} +
S_{\bar Q})\right],\label{1}\ee

\be S_0 = \frac{1}{4} \int d^4 x \left( F_{\mu\nu}^a\right)^2,\label{2}\ee \be
S_1 = -i \int d^4 x \bar \psi^f (\hat \partial +m )\psi^f,\label{3}\ee \be
S_{int} = - \int d^4 x \bar \psi^f g \hat A^a t^a \psi^f .\label{4}\ee

Here $f$ is flavor index, $S_Q$ and $ S_{\bar Q}$  refer to action of external
quark currents, of (possibly high  mass) quark $Q$ and antiquark $\bar Q$.

We exploit the background formalism \cite{9} to  split gluon field into
confining background $B_\mu$ and perturbative gluon field $a_\mu$ \be A_\mu =
B_\mu +a_\mu.\label{5}\ee

As in \cite{8} we shall use the simplest contour  gauge \cite{10} to express
$B_\mu$ in terms of field strength\footnote{ Since the whole construction of
$S_{eff}$ for quark $q$ in the field of antiquark $\bar Q$ is gauge invariant,
the final result does not depend on gauge \cite{11}, and the use of contour
gauge is a matter of convenience.}

\be  B_\mu (x) = \int_{C(x)} \alpha_\mu (u) F_{i\mu} (u) du_i, ~~\alpha_4 =1,~~
\alpha_i = \frac{u_i}{x_i}, \label{6}\ee and the contour $C(x)$ is going from
the point $x= (\vex, x_4)$ to the point $(\mathbf{0}, x_4)$ on the world-line
of $Q$ and then along this world-line to $x_4 =-\infty$. Note, that our final
result (\ref{11}), (\ref{12})  will be cast in the gauge invariant form, which
is the same  for all contours, connecting points $x,y$ to the world lines of
$Q$ (or $\bar Q$). The independence of the resulting asymptotic expressions
from the form of contours is shown in Appendix 3 of \cite{12}.

 Averaging over fields $B_\mu, (F_{\mu\nu})$, one can write
 \be
Z = \int \int D\psi D\bar\psi  \exp \left[ - (S_1 + S_{eff} )\right],\label{7}
\ee where $S_{eff}$ was computed in \cite{11}-\cite{12}. Keeping only quadratic
correlators  and colorelectric fields for simplicity, one obtains (for one
flavor)

\be
 S_{eff} = -\frac{1}{2 } \int d^4 x d^4 y
  \bar \psi   (x)\gamma_4[  \psi (x) \bar \psi (y)]\gamma_4  \psi (y)  J(x,y)
  \label{8}\ee  where  $[\psi \bar \psi]$ implies color singlet combination, and  $J(x,y)$  is expressed via vacuum correlator of
  colorelectric fields,
  \be  J(x,y) \equiv \frac{ g^2}{N_c}
  \lan A_4(x) A_4(y)\ran =  \int^x_0  du_i
  \int^y_0 dv_i D(u-v).\label{9}\ee
  Here $D(w)$ is the $np$ correlator, responsible for confinement \cite{13*},
  \be \frac{g^2tr}{N_c} \lan F_{i\mu} (u) F_{k\nu} (v) \ran = (\delta_{ik}
  \delta_{\mu\nu} - \delta_{i\nu} \delta_{\mu k} ) D (u-v) + O(D_1)\label{10}\ee
 and we have omitted the (vector) contribution of the correlator $D_1$,
containing perturbative  gluon exchange  and nonperturbative  ($np$)
corrections to it.

The properties of the kernel $J(x,y)$ have been studied in \cite{11,12},  here
we only mention the general form \be J(x,y) = \vex \vey f(\vex, \vey)
e^{(x_4-y_4)^2/4\lambda^2} D(0),\label{11}\ee where we assumed the Gaussian
form for simplicity $D(x) = D(0) e^{-\frac{x^2}{4\lambda^2}}$, and \be
f(\vex,\vey) = \int^1_0 ds \int^1_0 dt e^{-(\hat x s -\hat y t)^2}, ~~\hat x,
\hat y = \frac{\vex}{2\lambda} , ~~ \frac{\vey}{2\lambda},\label{12}\ee at
small $\hat x, \hat y$, $ f(0,0) =1$, while asymptotically \be f(\vex, \vey)
\cong \frac{\sqrt{\pi}}{\max (|\hat x|, |\hat y|)}, ~~\cos \theta
=1,\label{13}\ee where $\theta$ is the angle between $\vex$ and $\vey$. Note
also, that  $D(0)$ and $\lambda$ are connected to string tension $\sigma$ \be
\sigma= 2 \pi \lambda^2 D(0) = \frac12 \int D(x) d^2 x.\label{14}\ee We now
turn to the effective action (\ref{8}), where we   write explicitly all flavor
and color indices. In the latter case one should carefully restore the gauge
invariant combinations, derived in \cite{11},  using parallel transporters
$\Phi(u,v)=P\exp (\int^v_u A_\mu dz_\mu)$ and we denote \be \bar \psi_{\bar a}
(x) \psi_{\bar a} (y) \equiv \bar \psi_a (x) \Phi_{ab} (x, \bar Q, y) \psi_b
(y)\label{15}\ee with \be \Phi_{ab} (x, \bar Q , y) = \Phi_{ac} (\vex,x_4;
\mathbf{0}, x_4) \Phi_{cd} (\mathbf{0}, x_4; \mathbf{0}, y_4) \Phi_{db} (
\mathbf{0}, y_4, \vey, y_4),\label{16}\ee where $\mathbf{0}$ is at the position
of $\bar Q$.
 Thus (\ref{8}) can be rewritten as
\be S_{eff} =-\frac12 \int d^4 x d^4 y \bar \psi^f_{\bar a} (x) \gamma_4
\psi^f_{\bar b} (x) \bar \psi^g_{\bar b} (y) \gamma_4 \psi^g_{\bar a} (y)
J(x,y).\label{17}\ee We take now into account, that $\lambda\approx 0.1$ fm
\cite{13}, \cite{13*} is much smaller, than all hadron scales, and one can
integrate in (\ref{17}) over $d(x_4-y_4)$, using the form (\ref{11}), yielding

\be S_{eff} \approx -\int d \left(\frac{x_4+y_4}{2}\right) d^3 \vex d^3 \vey
(\bar \psi^f_{\bar a} (x) \gamma_4\psi^f_{\bar b} (x) )(\bar \psi^g_{\bar b}
(y) \gamma_4 \psi^g_{\bar a} (y)) \sigma (\vex\vey) \bar f(\vex, \vey)
\label{18}\ee where we have  used (\ref{14}) and defined $\bar f(\vex, \vey)
=\frac{f(\vex, \vey)}{2\lambda \sqrt{\pi}}$, so that $\bar f (\vex, \vex) \cong
\frac{1}{|\vex|}$, at large $|\vex|$.

To proceed to the practical calculations with the realistic baryon wave
functions, it is convenient to go over from bispinor  to $2\times 2$ formalism,
as it was done in \cite{14} for  $q\bar q$ vertices, see Appendix 2 of
\cite{14}. [Note, that the relativistic formalism for the hadron decay,
developed in \cite{8}, \cite{15}, and adapted for the baryon-antibaryon case in
Appendix below, accounts for the full relativistic structure of hadrons, and is
exemplified  in the factor $\bar y_{123}$, which is  the ratio of  the vertex
$Z_n$  factors for all hadrons. Below we follow a much simpler derivation in
terms of $2\times 2$ formalism, exploited in \cite{14}.]

We  now  take into account as in Appendix 2 of \cite{14}, that each bispinor
$\psi $ of light quark in (\ref{18}) obeys the Dirac one-body equation $(\veal
\vep+\beta (m+U) ) \psi = (\varepsilon -V)\psi$, where $U$ is the scalar
confining interaction, $U(x) =\sigma |\vex-\vex_{\bar Q}|$, and $V$ corresponds
to  perturbative gluon exchanges; therefore one can write
$\psi=\left(\begin{array}{l} v\\w\end{array}\right)$, where  \be
w=\frac{1}{m+U-V+\varepsilon} (\vesig \vep) v\to \frac{1}{m+\lan
U-V+\varepsilon\ran} \vesig \vep v\label{19}\ee where angular brackets imply
averaged value for a given quark in the given hadron, in our case this  refers
to  the average energy and potentials of a light quark in the  produced
heavy-light baryon , e.g. $\Lambda_c$. We also introduce for antiquarks
bispinors $\psi^c$ and spinors $v^c, w^c, \psi^c= (v^c, w^c)$. Therefore
$$\bar \psi =C^{-1} \psi^c = \psi^c (C^{-1})^T = \psi^c \gamma_2 \gamma_4;
\gamma_i =-i\beta \alpha_i$$ and \be \bar \psi\gamma_4 \psi = -i (v^c, w^c)
\beta \left(\begin{array}{ll} 0&\sigma_2\\\sigma_2& 0\end{array}
\right)\left(\begin{array}{l} v\\w\end{array}\right)= -i (\tilde v^c, \tilde
w^c)\left(\begin{array}{l}w\\v\end{array}\right)=-i (\tilde v^cw+ \tilde w^c
v)\label{20}\ee where notation is used, $v^c\sigma_2 \equiv \tilde v^c, ~~ w^c
\sigma_2 =- \tilde w^c= - \tilde v^c \vesig \overleftarrow{\vep}
\frac{1}{m+\lan U-V+\varepsilon\ran}.$ Hence (\ref{18}) can be written as (we
omit below  superscript $c$ in spinors $\tilde v^c$)\be S_{eff} = \int dt_4
d^3\vex d^3 \vey (\tilde v_{\bar a}^{fc} (\vex, t_4)K v^f_{\bar b} (\vex,
t_4))(\tilde v_{\bar b}^{g} (\vey, t_4)K v^g_{\bar a} (\vey, t_4))\sigma
(\vex\cdot\vey) \bar f (\vex, \vey)\label{21}\ee where \be K = \frac{1}{m+\lan
U-V+\varepsilon\ran} \vesig\vep +\vesig \overleftarrow{\vep}\frac{1}{m+\lan
U-V+\varepsilon\ran}\equiv \frac{\vesig (\vep+
\overleftarrow{\vep})}{\Omega}\equiv \frac{\vesig\veP}{\Omega}.\label{22}\ee
 We now form the $S$-wave
baryon wave function, which can be written as  a product of a symmetric
coordinate part and  antisymmetric spin-flavor-color factor
$A_{\mathcal{B}}$,\footnote{We neglect the nonsymmetric coordinate part of wave
function, which contributes less than one percent to the nucleon mass, see
\cite{17,16, 20} for more details. See also  \cite{20} for estimates of
$\Omega$ in (\ref{22}).}

 \be \Psi^{\mathcal{B}} =A_\mathcal{B} \Psi_\mathcal{B}^{(coord)}(\vex_1,\vex_2,\vex_3,
,t_4);~~A_{\mathcal{B}}=N_{\mathcal{B}}  \sum_{ijk} \frac{1}{\sqrt{6}} e_{abc}
C^{fgh}_{\alpha\beta\gamma}\varphi^f_{a\alpha} (i) \varphi^g_{b\beta}(j)
\varphi^h_{c\gamma} (k) \label{23}\ee where  $abc $ are color indices,
$\alpha\beta\gamma$ spinor indices and $fgh$ flavor indices.

One can separate the c.m. motion and define the  set   of  bound state wave
 functions  in the c.m. system $\{ \psi_h(\vex_1-\vex_3, \vex_2-\vex_3)\},$
 \be\Psi_\mathcal{B}^{(coord)}(\vex_1,\vex_2,\vex_3,
,t_4)=\frac{e^{-i  E t_4 - i
 \veP\veR}}{\sqrt{V_3}}  \Psi_{n} (\vexi,\veta),\label{24}\ee
 where  $\vexi, \veta$ are Jacobi coordinates, which can be defined in the relativistic case as
 \cite{17}, $(\omega_i = \lan \sqrt{\vep^2_i +m^2_i}\ran,~~ \omega_+= \sum_i
 \omega_i)$
 \be\veta = (\vex_2 -\vex_1) \sqrt{\frac{\omega_1\omega_2}{\omega_+
 (\omega_1+\omega_2)}},~~ \vexi = \sqrt{\frac{\omega_3}{\omega_+^2
 (\omega_1+\omega_2)}} (\omega_1\vex_1 +\omega_2\vex_2 - (\omega_1 +\omega_2)
 \vex_3)\label{25}\ee
 and $\Psi_n (\vexi, \veta)$ is expanded in the fast converging hyperspherical
 series, where the leading term ($>90\% $ in the wave function normalization,
 see \cite{17}, \cite{16} for details) is a function of hyperradius only, $\Psi_n (\vexi,
 \veta) \approx \psi (\rho)$, where
 \be \rho^2 = \sum^3_{i=1} \frac{ \omega_i}{\omega_+ }(\vex_i - \veR)^2 =
 \vexi^2+\veta^2.\label{26}\ee

 In what follows we shall be primarily interested in the   charmed baryons,
 $\Lambda_c, \sum_c,\Xi_c, \Omega_c$ and their orbital (and radial)
 excitations. As a first example we consider $\Lambda_c$ and take for
 simplicity only one (leading) component of wave function with singlet diquark
 made of $u,d$. The explicit forms of $A_{\mathcal{B}} $ for $p,\Lambda, \sum, \Xi$ are
 given in Appendix 1. For $\Lambda(\Lambda_c,\Lambda_b)$ one can   write in
 obvious notation

 \be A_{\Lambda_c}^{\alpha} = N_{\Lambda_c} \sum_{ijk} \frac{1}{\sqrt{6}} e_{abc}c_{a\alpha} (i)( (ud) - (du)) _{jk,bc}
 \label{27}\ee
 where $(ud)_{jkbc} \equiv  u_{\beta b} (j) d_{\beta c} (k) \varepsilon_{\beta}$,
 $\varepsilon_{\frac12}=- \varepsilon_{-\frac12} =1$.

 As shown in Appendix 2, the gauge invariant matrix element in the c.m. system
 of decaying charmonium   state $\Psi_{n_1} (\ver)$ can be written as
 \be G(n_1 \veP_1, n_2 \veP_2, n_3\veP_3)=(2\pi)^4 \delta^{(4)} (P_1-P_2-P_3)
 J^{\mathcal{B}\mathcal{\overline{B}}}_{n_1n_2n_3} (\vep)\label{28}\ee
 where $J^{\mathcal{B}\mathcal{\overline{B}}}_{n_1n_2n_3} (\vep)$ is
  \be J^{\mathcal{B\overline{ B}}}_{n_1n_2n_3} (\vep) = \int y_{123} e^{i\vep\ver} d^3 (\vex-\veu) d^3 (\veu-\vev) d^3
  (\vex-\vey) (\Psi_{n_1} \bar \mathcal{M} \Psi_{n_2} \Psi_{n_3}).\label{29}\ee

  Here $\ver = c (\veu-\vev),
  ~~c=\frac{\omega_Q}{\omega_Q+\omega_u+\omega_d},$
 and $\Psi_{n_i}$ are coordinate space spinor wave functions,while $\bar \mathcal{M}$ is defined as  \be \bar
 \mathcal{M}=\sigma (\vex\vey)\bar f(\vex, \vey) K_x K_y\label{30}\ee

At this point one needs to calculate the matrix element of the operator
$K_xK_y$ between $\mathcal{B}\bar{\mathcal{B}}$ wavefunctions, which we write
as \be \lan A_{\bar{\mathcal{B}}} | K_x K_y (\tilde v_{\bar Q}   \sigma_i v_Q)
| A_{\mathcal{B}}\ran = \eta_{\mathcal{B}Q} \frac{(\veP_x \veP_y)}{\Omega^2}
(\tilde v_{\bar \Lambda} \sigma_i v_{\Lambda})\label{31a}\ee Explicit
calculation yields coefficients $\eta_{\mathcal{B}Q}$, given  in Appendix 1 for
$\Lambda_c, \sum, p, \Xi$.

It is more convenient to go over to momentum  space in $J_{n_1n_2n_3} (\vep)$,
and using Appendix 2, Eq. (\ref{A2.15}), one has (we omit the superscript red
in (\ref{A2.15}) here and in  what follows) $$
J^{\mathcal{B}\mathcal{\overline{B}}}_{n_1n_2n_3} (\vep) = \int \bar y_{123}
\frac{d^3p_x}{(2\pi)^3}\frac{d^3p_y}{(2\pi)^3} \frac{d^3q_x}{(2\pi)^3}
\frac{d^3q_y}{(2\pi)^3} \Psi^+_{n_1} (c\vep -\vep_x -\vep_y)\times$$ \be \times
\Psi_{n_2} (\vep_x, \vep_y) \Psi_{n_3} (\vep_x + \veq_x, \vep_y
+\veq_y)\label{31}\ee where \be \bar y_{123} = \frac{ \veq_x \veq_y (\tilde
v_{\bar \Lambda} \sigma_i v_\Lambda)}{2\sqrt{2}N_c \Omega_x\Omega_y} \tilde
\mathcal{M} (\veq_x, \veq_y)\eta_{Q\Lambda}\label{32}\ee
  and $\tilde \mathcal{M}(\veq_x, \veq_y)$ is the Fourier transform of $\bar
  \mathcal{M}(\vex, \vey)$,  Eq. (\ref{30}), modulo  $K_x K_y$, the latter were
  taken into account  in the  prefactor of $\tilde \mathcal{M}$ in (\ref{32}).
  Also $\tilde
v_{\bar \Lambda}$ and $ v_{\Lambda}$ are spinors for $\Lambda^-_c$ and
$\Lambda^+_c$
  respectively, while $\sigma_i$ refers to the spin of $1^{--}(Q\bar Q)_n$
  state.
  From (\ref{A2.16}) one can write
  \be \tilde \mathcal{M} (\veq_x, \veq_y) =- \frac{\partial}{\partial \veq_x}
  \frac{\partial}{\partial\veq_y} \sigma\pi4\lambda^2 \int^1_0 \int^1_0  ds dt
  (2\pi)^3 \delta^{(3)} (t\veq_x + s\veq_y) e^\frac{-\lambda^2 (\veq_x-
  \veq_y)^2}{(s+t)^2}.\label{33}\ee
  Insertion of (\ref{33}) into (\ref{32}) and (\ref{31}) yields finally

$$ J^{\mathcal{B}\mathcal{\overline{B}}}_{n_1n_2n_3} (\vep) = \bar
y\int^1_0 \int^1_0dsdt \int  \frac{d^3p_x}{(2\pi)^3}\frac{d^3p_y}{(2\pi)^3}
\frac{d^3Q}{(2\pi)^3} e^{-\lambda^2\veQ^2 }\Psi_1^+ (c\vep -\vep_x- \vep_y)
\Psi_{2} (\vep_x, \vep_y)\times $$\be  \Psi_{3} (\vep_x +  s \veQ, \vep_y -
t\veQ),\label{34}\ee where we have differentiated by parts in $(\veq_x
\veq_y)\tilde M (\veq_x, \veq_y)$, obtaining \be \bar y = 4 \frac{3\cdot
2^{1/2} \pi}{N_c} \left(\frac{\sigma}{\Omega_u\Omega_d}\right) (\tilde v_{\bar
\Lambda}\sigma_i v_{ \Lambda})\eta_{Q\Lambda}.\label{35}\ee

Note the factor  4 in (\ref{35}), which comes from the  accounting for two
diagrams in Fig.1, and two diagrams with interchanging $u$- and $d$- vertices
between points $\vex$ and $\vey$.

\section{Baryonic width of heavy quarkonium and  the
$\mathcal{B}\bar{\mathcal{B}}$ yield in $\mathbf{e^+ e^-}$ collisions}

Using $J_{n_1n_2n_3} (\vep)$ in (\ref{34}), one can find the decay probability
of the $n_1$ state of $Q\bar Q$  into $\mathcal{B}\bar{\mathcal{B}}$ in the
states $n_2, n_3$ respectively, \be dw(1\to 23) =
\overline{|J^{\mathcal{B}\mathcal{\overline{B}}}_{n_1n_2n_3} (\vep)|^2}
(2\pi)^4 \delta^{(4)} (\mathcal{P}_1-\mathcal{P}_2-\mathcal{P}_3) \frac{d^3P_2
d^3P_3}{(2\pi)^6}\label{36}\ee where bar over $|J(\vep)|^2$ implies averaging
over initial and sum over final spin projections; in our simple case
$\overline{|\tilde v_{\bar \Lambda} \sigma_i  v_{\Lambda}|^2 } =1$. We now turn
to a more direct experimental process of $\mathcal{B }\bar{\mathcal{B}}$
production, namely $e^+e^-\to \mathcal{B}\bar{\mathcal{B}}$, which was observed
in \cite{3,4}. The corresponding amplitude can be written as \cite{18}

\be A_{n_2n_3} (\vep, E) = \sum_n c_n (E) T_{nn_2n_3}\equiv \sum_{n,m} c_n (E)
\left(\frac{1}{\hat E- E + \hat w(E)}\right)_{nm}
J^{\mathcal{B}\mathcal{\overline{B}}}_{mn_2n_3} (\vep)\label{37}\ee

Here $\hat E$ and $\hat w$  are matrices in indices $n,m,$ of the $Q\bar Q$
system, $ (\hat E)_{nm} = E_n \delta_{nm},$ \be w_{nm} (E) =\int
\frac{d^3p}{(2\pi)^3} \sum_{n_2n_3} \frac{J_{nn_2n_3} (\vep)
J_{mn_2n_3}(\vep)}{E-E_{n_2n_3}(\vep)},\label{38}\ee  where $n,m$ refer to the
complete set of charmonium bound states, and $J_{nn_2n_3} (\vep)$ is overlap
integral of the $n$-th charmonium state and $n_2, n_3$ states of heavy-light
mesons. In terms of $A_{n_2n_3}$ the total crossection is \be \sigma_{n_2n_3}
(E) = \int |A_{n_2n_3} (\vep, E)|^2 \pi \frac{d^3\vep}{(2\pi)^3} \delta
(E-E_{n_2n_3}(\vep))\label{39}\ee

The  factor $c_n(E)$ in (\ref{37}) accounts for the production of $(Q\bar Q)_n$
pair in the given process, in case of $e^+e^- \to (Q\bar Q)_n$ one has
\cite{18}
$$c_n
=\frac{4\pi\alpha e_Q\sqrt{6}}{E^2} \psi_n (0),$$ and with the definition
(\ref{37})

\be \Delta R_{n_2n_3} (E)= \frac{6\pi \cdot 12 e^2_Q}{E^2}  \sum |\psi_n (0)
T_{nn_2n_3} (E)|^2 d \Gamma_{n_2n_3},\label{40}\ee
where \be  d\Gamma_{n_2n_3}
=\pi \frac{d^3\vep}{(2\pi)^3} \delta (E-E_{n_2n_3}(\vep)).\label{41}\ee
As a result, keeping only one state $n$ in  (\ref{40}) one has  for
 \be
\Delta R_{n_2n_3} (E) = \frac{9 e^2_Q p(E) \psi^2_n(0)}{E }\frac{|J^{B\bar
B}_{n_1n_2n_3} (\vep) |^2}{|E_n-E+w_{nn}(E)|^2}\label{42}\ee where $J^{B\bar
B}_{n_1n_2n_3} (\vep)$ according to (\ref{A2.26}) can be written as

\be J^{B\bar B}_{n_1n_2n_3} = 2^{5/2}\pi^{1/4}
\frac{\sigma}{\Omega^2}\frac{\lambda^2\beta_0^{3/2} \mathcal{R}_n
(p)e^{-c\vep^2 R_0^2 \bar \Upsilon}}{\left( \frac{\lambda^2}{R^2_0}+ \bar
C\right)^{3/2} (1+ 2 \beta^2_0 R^2_0)^{3/2}},\label{43}\ee where parameters
$\beta_0, R_0, \bar\Upsilon, \bar C$ refer to $(Q\bar Q)_n$ and $
{\mathcal{B}\mathcal{\overline{B}}}$ wave functions and are defined numerically
in Appendix 2.

The polynomial $\mathcal{R}_n(p)$  is due to $(Q\bar Q)_n$ SHO wave function,
and  is obtained in the way described in Eq. (A.33). It can be approximated as
\be \mathcal{R}_n (p) \cong - 2.1 \left(1-0.034 \frac{p^2}{\beta^2_0} - 0.05
\left(\frac{p^2}{\beta^2_0}\right)^2\right)\label{44b}\ee

 In (\ref{43}) $\bar C$ and $\bar \Upsilon$ are values of $C$ and $\Upsilon$,
(\ref{A2.27}),(\ref{A2.28}) averaged over ($s,t$) integration  region.

A rough estimate of $\Delta R^{(n)}_{\mathcal{B}\bar\mathcal{B}}$ in
(\ref{42}), using (\ref{44b}), near $\Lambda^+_c \Lambda_c^-$ threshold  with
$\psi(4S)$ state for $(Q\bar Q)_n$  is \be \Delta
R^{(4)}_{\mathcal{B}\bar\mathcal{B}} \approx \xi \frac{p}{E} \frac{\exp (-2.5
\vep^2)}{|E-E_4+w_{44} (E)|^2},\label{45a}\ee with
\be \xi\cong0.9\cdot 10^4 e^2_Q \psi^2_4(0) \left(
\frac{\sigma}{\Omega^2}\right)^2 \frac{\beta^3_0
\lambda^4}{\left(\frac{\lambda^2}{R^2_0} + \bar C\right)^3 (1+2\beta^2_0
R^2_0)^3}\label{46a}\ee

Taking the Breit-Wigner form near the mass  $E_4$ of $\psi(4S)$, one can write
the cross section of $\sigma(e^+e^-\to \Lambda^+_c\Lambda^-_c)$ \be \sigma_4
(e^+e^-\to \Lambda_c^+\Lambda^-_c)=\xi~ 2.2\cdot 10^{-4} \frac{p}{E^3}
\frac{\exp(-2.5 p^2)}{(E-E_4)^2 + \frac{\Gamma^2_4}{4}} (mb),\label{47a}\ee
where all energies are in GeV and $\sigma_4$  in $mb$.

To calculate $\xi$ in (\ref{46a}), one can use
  \cite{21,22} average energies in (A.25) with  $\omega_n \ll \omega_c$, hence $a\approx 1,
b\approx 0$ and average values of  $s,t, \bar s = \bar t \approx 0.5$. This
yields $\bar C\cong 0.375, \bar \Upsilon\cong 0.3$.

Here  $E^2 =4 (\vep^2 + M^2_{\Lambda_c})$, and all momenta and energies are in
GeV. One expects, that $\lambda =O(1$ GeV$^{-1})$, as follows from  the
exponential fall-off of $D(x)$ in  \cite{13}, this value of $\lambda$ can be
varied for the Gaussian form used above. The values of
$\Omega_n=\frac{1}{m_n+\lan U-V\ran+\varepsilon_n}, n=u,d$, where averaging is
done over total baryonic state, and $m_u, m_d\approx 0$, can be found from the
analysis of baryons in \cite{20}, where for the light quark in a single orbital
with $\sigma =0.15$ GeV$^2$ one has $\varepsilon_{u,d} \approx 380$ MeV, while
$\lan U-V\ran$ can be roughly estimated as $0.6\div 0.8$ GeV, which yields
$\Omega_n \approx \Omega_d\approx (1\div 1.2)$ GeV. Now for the estimate of
$\psi_n(0)$ one can use calculations in \cite{21}, checked $vs$ experiment,
which give values of $R_{ns}(0)=\sqrt{4\pi} \psi_{ns} (0)$   with account of
mixing with $(n-1)^3D_1$ states.  Masses $E_n (~^3 S_1)$, given in Table below,
are calculated in \cite{21} (upper line) using flattening, and without
flattening in \cite{22} (lower line).

\vspace{1cm} {\bf Table}\\

\begin{tabular}{|l|l|l|l|l|l|l|}

\hline n&1&2&3&4&5&6\\\hline $E_n$, GeV& 3095& 3.682& 4.096& 4.426&
4.672&4.828\\
& (3.068)& (3.663)& (4.099)& (4.464)& (4.792)& (5.087)\\\hline $R_n (0)$,
GeV$^{3/2}$ & 0.905& 0.735& 0.511& 0.459& 0.360&$<0.445$\\\hline
\end{tabular}\\

\vspace{1cm}

Inserting $R_n(0) =0.46$ GeV$^{3/2}$  for $n=4$, from the Table  one obtains a
typical value of $\xi\approx 0.035 \lambda^4$, and the maximum of
$\sigma^{(4)}$ from (\ref{47a})  is of the order of 1 $nb$ for $\lambda\approx
1$ GeV$^{-1}$.
 This  magnitude is in accord with average experimental data in \cite{4}.

Note also, that $\exp (- 2.5 (\vep)^2 ) $ divided by the Breit-Wigner factor in
(\ref{47a}) is a strong cutoff factor, which decreases by a factor of 2 for
$\Delta E =0.2$ GeV from the threshold. As a result,  one obtains from
(\ref{47a}) the resonance-type
 behavior of $\sigma^{(4)}$ with maximum around $E=4.61$ GeV, and decreasing
 twice at E=4.7 GeV,  as shown in Fig.2. This form and the magnitude of the cross section
  correspond to experimental data in \cite{5}. Explicit calculations with realistic baryon wave
functions are  now in progress \cite{23}.

\begin{figure}[h]
\center{
\includegraphics[angle=0,width=0.55\textwidth]{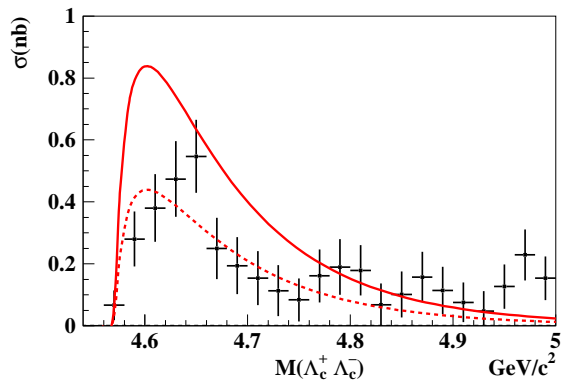}

\caption{ The cross section $ \sigma (e^+e^-\to \Lambda^+_c \Lambda^-_c)$ in
$nb$, estimated according  to Eq.(\ref{47a}) with  $\lambda=1$ GeV$^{-1}$ as a
function of total energy $E$ in GeV (solid line), experimental points are from
\cite{4}, dashed line is the  best normalization fit of Eq. (\ref{47a}) with a
factor of 0.52365.}}
\end{figure}

\section{Summary and discussion}

We have formulated above the fully nonperturbative mechanism for $
{\mathcal{B}\bar \mathcal{B}}$ production via double quark pair generation.
This mechanism is an extrapolation of the meson-meson production mechanism by
string breaking, studied recently in \cite{8}. Our main motivation is to
construct a nonperturbative theory of strong decays from the first principles,
in a similar way, as it was done in the theory of hadron spectra in one-channel
case, where all hadron masses are computed  from  the  first principle input:
 current quark masses, string tension $\sigma$ and $\alpha_s$, see \cite{21,
22} for recent results and references. The simple string-breaking mechanism was
indeed established  with the only  parameter $\sigma$   in \cite{8}, and
appeared to be
 close to the well-known phenomenological $~ ^3P_0$ model, in this way giving a theoretical foundation for the latter. In the present case
of ${\mathcal{B}\bar \mathcal{B}}$ production, an additional (fundamental)
parameter appeared: vacuum correlation length $\lambda$, which is connected to
the gluelump mass $\lambda=1/M_{gl}^{(2)}$ \cite{13}, and the latter is again
expressed via string tension $\sigma$: $M^{(2)}_{gl} = 6.15 \sqrt{\sigma}= 2.6$
GeV. In this way our first principle program is supported, however the
$O(\lambda^4)$ dependence of cross-sections makes the theory  very sensitive to
 a possible  process-depending renormalization of $\lambda$.

 It is clear, that the same mechanism should work for the pair  creation of
 other baryons,  containing $c(\bar c)$ quarks,   e.g. $\Sigma_c\bar \Sigma_c$,
 $\Xi_c\bar \Xi_c$, the only difference will be in coefficient $\xi$  and the
 dominant intermediate resonance $\psi(nS)$. In the general case one should sum
 up over $n$, as shown  in (\ref{37}) and a complicated  interference picture may
 appear.

 In this case, when only $\psi (4S)$ state was kept, and this state is not far
 from the $\Lambda^+_c\Lambda^-_c$ threshold, the resulting bump in Fig.2 is
 rather prominent. In Fig. 2  the predicted theoretical enhancement is compared  with experimental data from \cite{4}.
 One can see a reasonable agreement.

 The pair-creation mechanism, given in this paper, can be applied also to the
 case of light (strange) quarks $(Q\bar Q)$. In particular, for the  reaction $e^+e^-\to
 \Lambda\bar \Lambda, \Sigma\bar \Sigma$, studied experimentally  in \cite{26}, one can
 use the same equation (\ref{47a}), where the role of the intermediate state
 can play $\phi(2170)$ and higher $\phi$-mesons.

 Since radius of high-excited $\phi$'s is much  larger, than that of
 $\psi(4S)$, the corresponding $|\psi_n (0)|^2$ and $\beta_0$ in (\ref{46a})
 are smaller, and one expects the cros sections $\sigma (e^+e^-\to
 \Lambda\bar \Lambda, \Sigma\bar \Sigma)$, to be order of magnitude smaller than
 those for $
 \Lambda_c\bar \Lambda_c, \Sigma_c\bar \Sigma_c$. This is supported by
 experiments in \cite{26}.

 As for the case of the cross section $\sigma (e^+e^-\to
 \Lambda_b\bar \Lambda_b, \Sigma_b\bar \Sigma_b)$, our Eq. (\ref{47a}) applies
 here without modifications, except for the replacement of $\psi(4S)$   by $\Upsilon(6S)$; the main suppression factor
 comes from $E^3$ in the denominator of (\ref{47a}) and from $\left( \frac{e_b}{e_c}\right)^2 =\frac14$, while $|\psi_n(0)|^2$
  acquires factor 8.6, since $R_6(0) = 1.35
 $ GeV$^{3/2}$ \cite{27}. As a result the cross section  for $\Lambda_b \bar
 \Lambda_b$ production is one order of magnitude smaller than that for
 $\Lambda_c \bar \Lambda_c$ production.

One should stress, that the theory, developed here and in \cite{8}, can be
applied to string-breaking processes, where the energy  transfer $\Delta E$
from  ``external'' quarks $Q\bar Q$ to the pair-production vertex is not large,
$\Delta E \cdot \lambda\la 1$. In the opposite case one should take $\Delta E$
into account in the string profile function $J(x,y)$ in (\ref{11}), which
strongly changes   result, these effects are now  under investigation.

The theory,  proposed above, is  purely nonperturbative and therefore quite
different from the mostly perturbative  approach, developed before for
${\mathcal{B}\bar \mathcal{B}}$ production (see \cite{7}, \cite{27a} for
discussion and references). In this respect  two approaches complement each
other and the final goal  can be to formulate the unified theory,  where all
particle yields are expressed via first principle constants.

The  author is grateful  to  A.M.Badalian,  I.M.Narodetski,   and M.A.Trusov
for discussions and comments.  Useful advices, suggestions and help of
G.V.Pakh\-lova in preparing Fig.2 of the present paper are gratefully
acknowledged.The financial support of the Grant RFBR No 09-02-00620a is
acknowledged.

\newpage

\vspace{2cm}

{\bf Appendix 1  }

{\bf Baryon total wave function in terms of individual quark spinors

}

 \setcounter{equation}{0} \def\theequation{A1.\arabic{equation}}

\be A_{\mathcal{B}} = N_\mathcal{B} \sum_{abc} \frac{e_{abc}}{\sqrt{6}}
\sum_{ijk} C^\mathcal{B} (fgh |\alpha\beta\gamma) v^f_\alpha (a,i) v^g_\beta
(b,j) v^h_\gamma (c,k).\label{A1.1}\ee

We use notations  $v^u_{\frac12} (a,i) \equiv u_+ (a,i), ~~ v^d_{\frac12} (b,j)
=d_+ (b,j)$, etc. Here $i,j,k=1,2,3$ and $\sum_{ijk}$ denotes permutations of
1,2,3, we also require $j<k$, then the proton wave function with spin up can be
written as (color indices are suppressed for simplicity).

\be A_{{p}} = N_{p} \sum_{abc} \frac{e_{abc}}{\sqrt{6}} \sum_{ijk}
 u_+(i) [(d(j) u(k)) - (u(j) d(k))].\label{A1.2}\ee
 Here notation is used: $(du) = d_- u_+ - d_+ u_-; ~~ N_p
 =\frac{1}{3\sqrt{2}}$.

 For $\Lambda(J=\frac12)$ hyperon  with spin up one can write

 \be A_{{\Lambda}} = N_\Lambda \sum_{abc}
\frac{e_{abc}}{\sqrt{6}} \sum_{ijk} s_+(i)
[(u(j)d(k))-(d(j)u(k))],\label{A1.3}\ee

 and $N_\Lambda = \frac{1}{\sqrt{12}}$.

 For $\Sigma$ hyperons with spin up
 \be A_{\Sigma^0} = N_{\Sigma^0} \sum_{ijk} \sum_{abc}
 \frac{e_{abc}}{\sqrt{6}} \{ s_+ (i) ([ud]_0 + [du]_0) -2s_-(i) [ud]_{++}\},
 \label{A1.4}\ee
 where $ N_{\Sigma^0} =\frac16$ and  $[ud]_0 = u_+ (j) d_- (k) + u_- (j) d_+
 (k),$ $$[ud]_{++} = u_+ (j) d_+ (k) + d_+ (j) u_+ (k).$$

\be A_{\Sigma^+} = N_{\Sigma^+} \sum_{ijk} \sum_{abc}
 \frac{e_{abc}}{\sqrt{6}} \{ s_+ (i) ([udu]_0 - 2s_-(i)u_+ (j) u_+ (k) \},
 \label{A1.5}\ee
where $N_{\Sigma^+} =\sqrt{2}/6$.
 For ${\Sigma^-}$ one replaces in (\ref{A1.5}) all  $u$ quarks by $d$
 quarks. For $\Xi^0$ hyperon one has
$$A_{\Xi^0} = A_{\Sigma^+} (u\leftrightarrow s).$$

As a result of calculations of $\eta_{Q\mathcal{B}}$ one obtains \be
\eta_{s\Lambda} =1, ~~ \eta_{s\Sigma} = \frac19 = \eta_{u\Xi^0};~~\eta_{up}
=\frac43, ~~ \eta_{u\Lambda} = \frac{1}{3\sqrt{2}}.\label{A1.6}\ee

In the last two coefficients one must take into account the contribution of $u$
quark in the vector meson $(u\bar u)_n$, which gets into the  singlet pair
$(ud)$ or $(du)$. This contribution is proportional to $\tilde v_{\bar
\mathcal{B}} \vesig \veP v\cdot P_i \to \frac13\tilde v_{\bar \mathcal{B}}
\sigma_i v_{\bar \mathcal{B}}\veP^2$. In case of $\eta_{u\Lambda}$ the
isosinglet component of $(u\bar u)_n$ gives an extra factor of
$\frac{1}{\sqrt{2}}$.

 \vspace{2cm}

{\bf Appendix 2  }\\

{\bf Relativistic  derivation of the hadron $\to \mathcal{B   \overline{B}}$
amplitude }

 \setcounter{equation}{0} \def\theequation{A2.\arabic{equation}}

We start with the fully relativistic formalism and   we follow here the
derivation given in \cite{15}. The initial  stage is the point-to-point
amplitude, which is the Green's function $G_{123 xy}$ for $c\bar c$ state
emitted at point 1 and baryons absorbed at points 2 and 3, while intermediate
points $x,y$ are the same as in the main text, i.e. where two light quark pairs
of flavors $f$ and $g$ respectively are emitted, see Fig.1.

One can write this amplitude as

$$ \int G_{123xy} d^4x d^4y =\int d^4xd^4y tr (\Gamma_1 S_Q (1,2) \Gamma_2 S^f_q
(2,x) S^g_q (2,y)\times $$ \be\times \Gamma_x  \mathcal{M} (x,y) \Gamma_y S^f_q
(x,3) S_q^g (y,3) \Gamma_3 S_q(3,1)) \equiv \lan 0 |j_Q (1) j_\mathcal{B} (2)
\mathcal{M}j_{\bar\mathcal{ B}} (3)|0\ran \label{A2.1}\ee where $S_{q, Q}$ are
light $(q)$ and heavy $(Q)$ quark propagators, and $\Gamma_i$ are vertices for
given hadrons, e.g. $\Gamma_1 =\gamma_i$ for  $1^{--}$ state of charmonia etc.,
while $\Gamma_x =\Gamma_y =\gamma_4$. Finally, \be \mathcal{M} (x,y) = \sigma
(\vex \vey) \bar f (\vex, \vey)\label{A2.2}\ee and one  should integrate
(\ref{A2.1}) over $d^4 x d^4y$. However, the physical
 amplitude of  a hadron decay into  two hadrons $A(n_1 \veP_1; n_2 \veP_2, n_3
 \veP_3)$ should be obtained from $G_{123xy} $ in two steps: 1) first one
 should go from coordinate points 1,2,3 to definite momentum states $\veP_1,
 \veP_2, \veP_3$, and  2) one should go from point-to-point amplitude to
 hadron-to hadron amplitude, which is obtained by amputating in the matrix
 element (\ref{A2.1}) the pieces $\lan 0 | j_i | n_i \veP_i \ran $, which are
 proportional to hadron decay constant. E.g. for a vector meson
 \be \lan 0 | j^\Gamma_k |n, \veP =0 \ran = \varepsilon _k \sqrt{\frac{M_n}{2}
 }f^{(n)}_\Gamma\label{A2.3}\ee

 Proceeding as in Appendix 2 of  \cite{15}, one arrives at the expression
 \be A(n_1 \veP_1; n_2 \veP_2 , n_3\vep_3) = (2\pi)^4 \delta^{(4)}
 (\mathcal{P}_1 - \mathcal{P}_2 - \mathcal{P}_3)  J_{n_1n_2n_3}^{(rel)}
 (\vep)\label{A2.4}\ee
 where

 $$J^{rel}_{n_1n_2n_3} (\vep) = \frac{1}{N_c} \int   \bar y_{123}  d^3 (x-u) d^3
 (u-v)  d^3(x-y) \Psi_{n_1} (\veu -\vev) \sigma (\vex \vey) e^{i\vep\ver} \bar f( \vex \vey)\times$$
\be\times  \Psi_{n_2} (\vex-\veu, \vey-\veu) \Psi_{n_3} (\vex-\vev,
 \vey-\vev),\label{A2.5}\ee
 and $\ver =c (\veu-\vev), ~~c=\frac{\omega_c}{\omega_c+\omega_u+\omega_d}$,
 where  $\omega_i$ is average kinetic energy of quark $i$ in the hadron.
 Here $\Psi_{n_i}$ are coordinate parts of wave functions\footnote{We assume
 here for simplicity, that a relativistic state  can be described by only  one
 scalar
 function, otherwise one has to sum over all terms with coefficients $\bar
 y_{123}^{(i)}$, specific for each term $i$}, and $\bar y_{123}$ is computed as
 a ratio of total trace and hadron $Z_i$ factors, (see Appendix 2 of \cite{15} for
 details)
 \be \bar y_{123} = \frac{Z_{123 xy}}{\sqrt{Z_1Z_2Z_3}}, ~~ Z_1 = tr (\Gamma_1
 \Lambda_Q \Gamma_1 \Lambda_{\bar Q})\label{A2.6}\ee

 \be Z_k (k=2,3) = tr (\Gamma_k \prod^3_{s=1}\frac{(m-i\hat
 p_s)}{2\omega_s} \Gamma_k) \label{2.7}\ee

 \be Z_{123 xy} = tr(\Gamma_i \Lambda_Q \Gamma_2 ( \Lambda_{\bar q} \gamma_4
 \Lambda_q) (\Lambda_{\bar q} \gamma_4 \Lambda_q) \Gamma_3 \Lambda_{\bar
 Q})\label{A2.8}\ee
 and
 \be \Lambda_q =\frac{m_q-i p \gamma_i + \omega_q \gamma_4}{2\omega_q} , ~~
\Lambda_{\bar q}  =\frac{m_{\bar q}-i p \gamma_i - \omega_q
\gamma_4}{2\omega_q} .\label{A2.9}\ee Here $\omega_q =\lan
\sqrt{m^2_q+\vep^2}\ran$, where the average is for the given hadron $n$.

Examples of $\bar y_{123}$ for meson$\to $ 2 meson decay are given in
\cite{14,15}.

A much simpler derivation can be made in the so-called Dirac formalism,
introduced in \cite{14}. In this case the final expressions are given in the
form of $2\times 2$ matrices and it is   convenient  in this case to write in
(\ref{A2.4})  the Dirac-reduced expression   $J^{red}_{n_1n_2n_3}$ instead of
$J^{rel}_{n_1n_2 n_3}$, and the former is best written in the momentum space
(first in the simpler case, when $\mathcal{M} (x,y)$ in (\ref{A2.2}) is taken
as an effective constant $ \bar \mathcal{M}$). \be J^{'red}_{n_1n_2n_3} (\vep)
= \int y'_{red} \frac{d^3p_x}{(2\pi)^3} \frac{d^3p_y}{(2\pi)^3} \Psi^+_{n_1}
(c\mathbf{p}-\mathbf{p}_x-\mathbf{p}_y) \Psi_{n_2} (\mathbf{p}_x, \mathbf{p}_y)
\Psi_{n_3}(-\mathbf{p}_x, -\mathbf{p}_y)\label{A2.10}\ee where\be y'_{ red} =
tr \{ \Gamma^{(n_1)}_{red} \Gamma_{red}^{(n_2)} K(p_x) \bar \mathcal{M} K(p_y)
\Gamma^{(n_3)}_{red}\},\label{A2.11}\ee and $K$ defined in (\ref{22}).

Note, that (\ref{A2.11}) has the same structure, as Eq. ({B3}) in \cite{14},
with ($\vesig\veq)$ replaced by $K\mathbf{M}K$ in (\ref{A2.11}).

As follows from the Table VII in \cite{14}, the reduced vertex
$\Gamma_{red}^{(n_1)} =\frac{1}{\sqrt{2}} \sigma_i$ for $~^3S_1$ state
$(1^{--}$) of charmonium, while  for $\Gamma_{red}^{(n_k)}, k=2,3$, one must
choose the appropriate baryon vertex, which  is given in Appendix 1. To
illustrate our procedure, we consider a simplified example (where color indices
are  suppressed, but the final result coincides with the exact one) of a
baryon, consisting of singlet $(ud)$ pair plus $c$ quark, as \be
(\Gamma^{(n_2)}_{red})_{\alpha \beta \gamma} =\frac{
\varepsilon_{\alpha\beta}}{\sqrt{2}} \chi_\gamma;
(\Gamma^{(n_3)}_{red})_{\alpha' \beta' \gamma'} =
\frac{\varepsilon_{\alpha'\beta'} }{\sqrt{2}}\chi_{\gamma'}^+.\label{A2.11a}\ee
Now, introducing $\veP$ in $K(\veP)$, so that  $ K=\frac{\vesig \veP}{\Omega}$,
with $\veP=\vep + \overleftarrow{\vep}$,

 $$ \varepsilon K (\veP_x) K(\veP_y) \varepsilon
= \varepsilon_{\beta\gamma} K_{\beta\alpha} (\veP_x) K_{\gamma\delta} ( \veP_y)
\varepsilon_{\alpha \delta},$$ one obtains \be (\varepsilon KK \varepsilon)
=-\frac{ 2 \veP_x \veP_y }{\Omega_x\Omega_y} \label{A2.12}\ee and finally \be
y^{red}= -\frac{1}{\sqrt{2}} \frac{\veP_x \veP_y }{\Omega_x\Omega_y} (\chi^+
\vesig_i \chi)\bar \mathcal{M}\label{A2.13}\ee and for $\Omega_x\Omega_y$ can
be assigned the values $\Omega_u,\Omega_d$ (or vice versa)  with $\Omega_{u,d}
= {m_{u,d}+ \lan U-V\ran+ \varepsilon_{u,d}}$.

However, $\Omega_{u,d}$ in (\ref{A2.12}) can be easily extracted from $K$ in
Eq. (\ref{22}) of the main text, and one can see, that in the approximation,
when the denominator in $K$  is kept constant (independent of $\vex$ or $\vey$)
 $K_x \sim \vesig (\vep_x+ \vep'_x) = 0$.
Therefore one  must now take into account the coordinate dependence of $
\mathcal{M} (x,y)$ in (\ref{A2.2}) and we  write \be \mathcal{M} (x, y) =
\mathcal{M} (\vex-\veu, \vey-\veu) = \int\tilde \mathcal{M} (\veq_x, \veq_y)
\frac{d^3q_x}{(2\pi)^3} \frac{d^3q_x}{(2\pi)^3} e^{i\veq_x(\vex-\veu)+ i \veq_y
(\vey-\veu)}\label{A2.14}\ee and (\ref{A2.10}), (\ref{A2.11}) are replaced by
\be J^{red}_{n_1n_2n_3} (\vep) =\int \bar y^{red} \frac{d^3p_x}{(2\pi)^3}
\frac{d^3p_y}{(2\pi)^3} \frac{d^3q_x}{(2\pi)^3}
\frac{d^3q_y}{(2\pi)^3}\Psi^+_{n_1} (p-p_x-p_y)\times\label{A2.15}\ee
$$\times \Psi_{n_2} (p_x, p_y) \Psi_{n_3}(-p_x-q_x, -p_y-q_y) $$ where $y^{'red}$ in (\ref{A2.10}) is
replaced by $ \bar y^{red} = y^{red}\tilde\mathcal{M} (q_x, q_y)$, and
$y^{red}$ in (\ref{A2.13})

Here $ \mathcal{M}(q_x, q_y)$ is the Fourier transform of $ \mathcal{M} (x,y)$
(\ref{A2.2}),\be \mathcal{M} (q_x, q_y) =\int\int d^3 x d^3 y \frac{\sigma
(\vex\vey)}{2\lambda\sqrt{\pi}} \int^1_0\int^1_0 ds dt \exp \left[ -
\frac{(\vex s -\vey t)^2}{4\lambda^2}\right] e^{-i\veq_x\vex
-i\veq_y\vey}.\label{A2.16}\ee Performing the integrals, one obtains \be
\mathcal{M} (\veq_x, \veq_y) = -\frac{\partial}{\partial\veq_x}
\frac{\partial}{\partial\veq_y}\frac{\sigma
\pi^{3/2}}{2\lambda\sqrt{\pi}}(2\lambda)^3 \int^1_0\int^1_0 ds dt (2\pi)^3
\delta^{(3)} (t\veq_x + s\veq_y) e^{-\frac{\lambda^2 (\veq_x
-\veq_y)^2}{(s+t)^3}}.\label{A2.16}\ee Insertion of (\ref{A2.16}) into
(\ref{A2.15}) yields (after integrating out $\delta$- function in (\ref{A2.16})
and differentiating $\frac{\partial}{\partial \veq_x}~~
\frac{\partial}{\partial \veq_y}$ by parts)

\be J^{red}_{n_1n_2n_3}(\vep)\equiv J(\vep)= \int^1_0 \int^1_0 dsdt ~~\bar
y^{\rm red}\int \frac{d^3p_x}{(2\pi)^3} \frac{d^3p_y}{(2\pi)^3}
\frac{d^3Q}{(2\pi)^3}\Psi_1(c\vep-\vep_x-\vep_y)\times \label{A2.19}\ee
$$\Psi_2(\vep_x, \vep_y)\Psi_3(\vep_x+s\veQ; \vep_y-t\veQ),$$

where $\bar y^{\rm red}$ is now \be \bar y^{\rm red} =\frac{3\cdot 2^{1/2}
\lambda^2 \pi  \sigma (\chi^+ \sigma_i \chi)}{N_c \Omega_u\Omega_d}
e^{-\lambda^2 \veQ^2}.\label{A2.20}\ee

To estimate the integral in (\ref{A2.19}) we use the oscillator wave functions
for $\Psi_1$, and oscillator form of hyperspherical wave function for $\Psi_2,
\Psi_3$,

\be \Psi_1 (\vex) = \mathcal{P}(\vex) \exp \left( - \frac{\beta^2_0
x^2}{2}\right), ~~ \Psi_2 (\rho) = N\exp \left( -
\frac{\rho^2}{\rho^2_0}\right), \label{A2.21}\ee where $\rho^2 = \vexi^2 +
\veta^2= \sum^3_{i=1} (\vex^{(i)} - \veR)^2 \frac{\omega_i}{\omega_+} , ~~
\omega_i = \lan \sqrt{ p^2_i+m^2_i}\ran, ~~ \omega_+ =\sum^3_{i=1} \omega_i$
$\omega_u\cong \omega_d\equiv \omega_n.$

In $p$-space one can use (see (25)  and  \cite{17} for relations between
standard and Jacobi coordinates) \be \Psi_2 = N_2 \exp (- R^2_0
(\vep^2_\xi+\vep^2_\eta)), ~~ N^2_2 = (8\pi R^2_0)^3\label{A2.22}\ee and
$\vep_\xi = \sqrt{\frac{\omega_+}{2\omega_c}}(\vep_x +\vep_y), ~~ \vep_\eta
=\frac{1}{\sqrt{2}} (\vep_y -\vep_x)$.

Relation between average $\lan \rho^2\ran $ and $R^2_0$ is $\lan \rho^2\ran
=6R^2_0=\lan r^2_\mathcal{B}\ran $. For $\Psi_1$ one can  first take for
simplicity $\Psi_1(p) =N_1 \exp \left(-\frac{\vep^2}{2\beta^2_0}\right).$

Now one can integrate in(\ref{A2.19}) over $d^3p_x d^3p_y$, which yields
$$I(\vep , \veQ) \equiv \int \frac{ d^3p_x}{(2\pi)^3} \frac{d^3p_y}{(2\pi)^3}
\Psi_1(c\vep-\vep_x-\vep_y) \Psi_2(\vep_x, \vep_y)\Psi_3(\vep_x+s\veQ,
\vep_y-t\veQ)=$$ \be=N_1N_2^2 \left( \frac{1}{\sqrt{d_1(a-b)
R^2_0}8\pi}\right)^3 \exp (-\Xi),\label{A2.23}\ee

where \be \Xi= \frac{(c\vep)^2}{2\beta^2_0} + a R^2_0 \veQ^2 (s^2 +t^2) - 2 b
R^2_0 \veQ^2 st - \frac{\mathbf{d}^2_2}{4d_1} - \frac{(s+t)^2}{4} \veQ^2 R^2_0
(a-b)\label{A2.24}\ee and \be a= \frac{ (\omega_++\omega_c)}{2\omega_c}, ~~ b=
\frac{\omega_n}{\omega_c}, ~~d_1 =\frac{1}{2\beta^2_0} + R^2_0 (a+b) ; ~~
\mathbf{d}_2 =-\frac{c\vep}{\beta^2_0}+R^2_0 (a+b)\veQ(s-t).\label{A2.25}\ee

Finally the integration over $d^3Q$ can  be done in (\ref{A2.19}), yielding \be
J(\vep) =\tilde y^{\rm red} \frac{N_1N_2^2}{(4\pi)^6} \int^1_0\int^1_0  ds dt
\left[ \frac{\pi}{(\lambda^2 + C R^2_0) d_1 R_0^2 (a-b)} \right]^{3/2} \exp
[-R^2_0 ( c\vep)^2 \Upsilon]\label{A2.26}\ee

where
 \be \Upsilon = \frac{a+b}{1+2\beta^2_0 R^2_0 (a+b)}
 - \left(\frac{a+b}{1+2\beta^2_0 R^2_0 (a+b)}\right )^2
  \frac{R^2_0}{4(\lambda^2 + C
R^2_0)}, \label{A2.27}\ee
 and \be C= a(s^2+t^2) - 2 bst - \frac{(s+t)^2}{4}
(a-b) - \frac{(a+b)^2 (s-t)^2 R^2_0\beta^2_0}{2(1+2\beta^2_0 R^2_0 (a+b))},
\label{A2.28}\ee

 \be \tilde y^{red} =
\frac{3\cdot 2^{3/2}\pi  \lambda^2 \sigma (\chi^+ \sigma_i \chi)}{
N_c\Omega_u\Omega_d}\label{A2.29}\ee

In (\ref{A2.26}) $N_1=\left(\frac{8\pi}{2\beta^2_0}\right)^{3/4}$, if the SHO
for $\psi_1$ is used.

For $(n^3S_1)c\bar c$ state in the same oscillator basis one should use
instead, as in \cite{14}\be \Psi_1(n^3S_1)= \frac{1}{\sqrt{4\pi}}
R^{SHO}_n(\beta_0, p) = \frac{(-)^n (2\pi)^{3/2}}{\sqrt{4\pi} \beta_0^{3/2}}
\sqrt{\frac{2(n-1)!}{\Gamma(n+\frac12)}} e^{-\frac{p^2}{2\beta^2_0}
 }L^{1/2}_{n-1}\left(\frac{p^2}{\beta^2_0}\right)\label{A2.30}\ee normalized as
$\int |\Psi_1|^2 \frac{d^3p}{(2\pi)^3}=1$.

To understand the structure of the obtained result (\ref{A2.25}), one can use
the limit of large mass $m_Q$, i.e. $\omega_c \gg\omega_n$, $n=u,d$,  which
yields $a=1, b=0$. Another useful limit is a) $2 \beta^2_0 R^2_0\gg 1$, which
is achieved for large size   baryons as compared to the  radius of charmonium:
note, that  $R^2_0 \cong \frac16\lan r^2_B\ran$,   while $\beta_0 (2S) = 0.46$
GeV \cite{14}. In this case one obtains \be C=\frac34 (s^2+t^2), ~~ \Upsilon =
1- \frac{1}{16 \beta_0^4 R^2_0 (\lambda^2+ CR^2_0)}.\label{A2.31}\ee

In the opposite limit: b) $2 \beta^2_0 R^2_0 \ll 1$ one has \be C=s^2+t^2 -
\frac{(s+t)^2}{4} - \frac{(s-t)^2}{2} R^2_0 \beta^2_0, ~~ \Upsilon =  1 -
\frac{R^2_0}{4(\lambda^2+ CR^2_0)}.\label{A2.32}\ee

In both cases one can use our result (\ref{A2.24}) for the simple Gaussian form
of $\Psi_1$ to derive the final result for a more  complicated function
(\ref{A2.30}), by using \be p^2 e^{-\frac{p^2}{2\beta^2}} =-
\frac{\partial}{\partial(1/ (2\beta^2_0))}
e^{-\frac{p^2}{2\beta^2_0}},\label{A2.33}\ee or directly, introducing in
(\ref{A2.23}) the $\Psi_1$ given in (\ref{A2.30}).

With the simple exponential function $\Psi_1$  one has   $N_1=
\left(\frac{4\pi}{\beta^2_0}\right)^{3/4}$   and for $ R_0^2\beta^2_0\ll 1 $
one has an estimate \be J(\vep) = \tilde y^{\rm
red}\left(\frac{\beta^2_0}{\pi}\right)^{3/4} \frac{\exp (-0.6~
R^2_0\vep^2)}{\left(\frac{\lambda^2}{R^2_0}+C\right)^{3/2}} .\label{A2.34}\ee
Finally, one should take into account also, that $\tilde y^{\rm red}\equiv
\tilde y^{\rm red}_{du} (\bar Q)$, while the total coefficient should be

 \be \tilde y^{\rm
red}_{\rm total} = \tilde y^{\rm red}_{ud} (\bar Q)+\tilde y^{\rm red}_{du}
(\bar Q)+\tilde y^{\rm red}_{ud} ( Q)+\tilde y^{\rm red}_{du} ( Q)= 4 \tilde
y^{\rm red}_{du} \label{A2.35}\ee and this is the value, which should be
introduced in (\ref{A2.25}) instead of $ y^{\rm red},  ~~ y^{\rm red}\to\tilde
y^{\rm red}_{\rm total}$.
\end{document}